\documentclass[preprint2]{aastex}

\usepackage{natbib}           % Manuel: attached, not original in this file
\usepackage{amsmath}
\usepackage{chngcntr}
\usepackage{multicol}
\usepackage{appendix}
\usepackage{graphicx}       % Manuel: attached, not original in this file
 \usepackage{epstopdf}      % Manuel: attached, not original in this file
\shorttitle{\ion{Mg}{2} h+k - Rotational Period in G-type stars}
\shortauthors{Olmedo et al.}

\begin{document}


\title{\ion{Mg}{2} h+k Flux - Rotational Period Correlation for G-type stars}


\author{Manuel Olmedo, Miguel Ch\'avez, Emanuele Bertone, V\'ictor De la Luz}
\affil{Instituto Nacional de Astrof\'isica Optica y Electr\'onica \\
    Luis Enrique Erro \#1 C.P. 72840, Tonatzintla, Puebla, M\'exico}

\email{olmedo@inaoep.mx, mchavez@inaoep.mx, ebertone@inaoep.mx, itztli@gmail.com}


%\altaffiltext{1}{Visiting Astronomer, Cerro Tololo Inter-American Observatory.
%CTIO is operated by AURA, Inc.\ under contract to the National Science
%\altaffiltext{2}{Society of Fellows, Harvard University.}
%\altaffiltext{3}{present address: Center for Astrophysics,
%    60 Garden Street, Cambridge, MA 02138}
%\altaffiltext{4}{Visiting Programmer, Space Telescope Science Institute}
%\altaffiltext{5}{Patron, Alonso's Bar and Grill}


\begin{abstract}
We present an analysis of the correlation between the mid-UV \ion{Mg}{2} h and k emission lines and measured rotational 
periods of G-type stars. Based on IUE and HST high resolution spectra of a sample of 36 stars, we derive 
an exponential function that best represents the correlation. We find that the variation of the \ion{Mg}{2} 
${\rm{h+k}}$ fluxes is about a factor of 2.5 larger than that of \ion{Ca}{2} H+K, indicating that the UV 
features are more sensitive to the decline of $P_{\rm{rot}}$. The comparison of UV-predicted rotational 
periods with those derived from empirical $P_{\rm{rot}}$ - \ion{Ca}{2} H+K flux calibrations are consistent, 
with some scatter at large periods, where the emission are less intense. We present newly derived rotational periods 
for 15 G-type stars.
\end{abstract}

 
\section{Introduction} % ==============================================================

The rate at which main sequence stars rotate is certainly amongst the most fundamental  stellar parameters. 
It has proven to be a fundamental tool in a variety of stellar astrophysical phenomena, in particular as a 
test of the dynamo theories. In the time domain, rotational periods are fundamental for understanding the 
evolution of angular momentum in low mass stars, from the T-Tauri stage to mature objects as our Sun. 
Additionally, rotational velocity has become a leading tool for determining stellar ages of main sequence
stars in the field, for which other methods, such as the isochrone fitting, are difficult to apply 
\citep{2010ARA&A..48..581S}.

Direct measurements of the stellar rotational periods rely on the analysis of brightness variations on the stellar surface due to the presence of spots. Nevertheless, aside from highly magnetic objects of early-type where the spot coverage can be large, this method represents, in many instances, a difficult task for cooler stars. In G-type stars the flux fluctuation is rather weak and requires very long and precise photometric and spectroscopic observational programs. In order to cope with these potential constraints, other indirect methods have been implemented, mainly through the analysis of the rotationally driven chromospheric activity.

The correlation between the stellar chromospheric activity and rotation was first identified 
in the pioneering work of \cite{1967ApJ...150..551K}. This relationship is generally explained by 
the balance between an enhanced magnetic field induced by rotation and the breaking due to loss of 
angular momentum by magnetized stellar winds \citep{1968MNRAS.138..359M}. An additional part of the 
picture is given by the internal magnetic fields  generated in the interiors of convective stars 
($T_{\rm eff}<8000\phantom{a}\rm K $) through a solar-like dynamo mechanism.

Magnetic activity has been usually measured by the emission excess in the strong \ion{Ca}{2} 
H and K features. The cardinal work initiated at Mount Wilson in the early sixties 
\citep{1963ApJ...138..832W} was later continued through a number of extensive surveys 
\citep[see, for instance,][]{1991ApJS...76..383D,1996AJ....111..439H,2004ApJS..152..261W,2008A&A...485..571J,2013AJ....145..140Z}

Whilst the \ion{Ca}{2} H and K lines have long been the workhorse for stellar activity  studies, in particular
due to their accessibility from ground based telescopes, the \ion{Mg}{2} h and k emission lines in 
the mid-UV (2800\phantom{a}\AA) has provided an additional valuable diagnostics. These lines, with 
a similar energetics as those of \ion{Ca}{2}, present the advantage that their shallower absorption 
wings make the emission line easier to measure and less contaminated by the photosphere.

Early works on the \ion{Mg}{2} h and k resonance lines by \cite{1984ApJ...279..778H} demonstrated 
that this chromospheric proxy strongly correlates with rotational period, $P_{\rm{rot}}$. Their 
study was based on a relatively small sample (31 objects) covering a wide range of mass and 
evolutionary status, and also included numerous binaries. More recently, \citet[][hereafter 
CC07]{2007ApJ...666..393C} analyzed  a larger stellar sample and identified a loose correlation 
between \ion{Mg}{2} flux and rotational period, a situation that is improved if age instead of 
rotational period is used.

In this paper, we revisit the \ion{Mg}{2} h and k doublet as a powerful diagnostics of the 
rotational period of G-type stars by analyzing a stellar sample of main sequence and subgiant 
stars observed by International Ultraviolet Explorer (IUE), complemented with high resolution 
observations conducted by the Space Telescope Imaging Spectrograph (STIS) and the Goddard 
High Resolution Spectra (GHRS) on board the Hubble Space Telescope (HST). We calibrate 
the \ion{Mg}{2} h and k absolute fluxes with measured  rotational periods and apply the 
relation  to determine new rotational periods for 15 stars with high quality UV spectra. 

\section{The Stellar Sample}  % ==============================================================

 % \footnote{http://sdc.cab.inta-csic.es/ines/index_esp.html}

The stellar sample upon which our analysis is based has been constructed from two different databases.
The first is that of the IUE-Newly Extracted Spectra (INES\footnote{http://ines.ts.astro.it/cgi-ines/IUEdbsMY}):
we searched for all high resolution data available for objects in class 44 (G-type stars of luminosity 
classes IV and V). This search delivered nearly 500 spectra that were subsequently visually inspected to 
verify their quality. A total of 57 objects have been selected from this database. We have also searched 
available data of G-type stars from HST and found 19 objects, 17 are from STIS, 1 from GHRS, and 1 object have 1 image from both instruments; these data were taken from the HST-MAST archive\footnote{http://archive.stsci.edu/hst/} and  StarCat \citep{2010ApJS..187..149A}. For the solar spectra, we extracted a high resolution spectrum of the Moon listed in class 02 of the INES archive.

% The error in the measurements

The full sample consists of 75 object plus the Sun.  The stellar set is divided into three working sub-samples. The first sub-sample consist of 37 stars (and the Sun) with rotational periods, that we call {\it primary}, and will be used to establish the calibration of the \ion{Mg}{2} UV fluxes vs.~rotational period. 
For these objects, the rotational periods have been measured either through the analysis of the photometric modulation of the visible flux due to the uneven distribution of stellar spots on the stellar surface or through variability of the \ion{Ca}{2} emission. The data for these stars are given in Table 1 where columns 1-7 provide, respectively, the star name, the spectral type, the color index $B-V$, the effective temperature mainly obtained from the compilation of \cite{2010A&A...515A.111S}, the calculated \ion{Mg}{2} absolute flux (see next section), the rotational period, and a label of the reference for the period. The list of reference is given at the bottom of the table.
The second sub-sample is composed of 23 objects with {\it secondary} rotational periods, this is, periods determined through a calibration of \ion{Ca}{2} emission and measured rotational periods. This dataset will serve as a test to compare rotational periods derived with our UV calibration with those computed from \ion{Ca}{2}. Table 2 lists the objects included in this set. Column 1-5 are as in Table 1, while the last 3 columns provide the secondary rotational period, the reference for this period, and the rotational period estimated in this work, respectively.
Finally, the third collection corresponds to 15 stars with no available information on their rotational periods, and for which we derive the first estimate of this parameter. This latter set is presented in Table 3, where we also provide our rotational periods and their estimated errors.

\begin{table}[ht!]
\caption{Basic data for stars with primary rotational periods}
\vspace{-12pt}
\begin{center}
\footnotesize
\tabcolsep=0.08 cm
\begin{tabular}{llllcll}

\hline \hline \rule{0pt}{2.5ex} 
Star & Spectral & $B-V$ & $T_{\rm{eff}}$ & $\log F_{\rm{h+k}}$ & $P_{\rm{rot}}$ & Ref. \\
&Type&&\scriptsize{[K]}&&\scriptsize{[days]}&\\ [0.5ex]
 \hline \rule{0pt}{2.5ex} 
  Sun    &    G2V    &    0.65    &    5775    &    6.02    &    26.09    &    a    \\
  HD1835		&    G3V    &    0.67    &    5776    &    6.47    &    7.78    &    a    \\
  HD9562		&    G1V    &    0.603    &    5849    &    5.68    &    29    &    b    \\
  HD10700    &    G8.5V    &    0.72    &    5328    &    5.72    &    34    &    b    \\
  HD11131    &    G1Vk:    &    0.632    &    5767    &    6.47    &    8.92    &    z    \\
  HD13974    &    G0V    &    0.58    &    5606    &    6.29    &    11.1    &    l    \\
  HD20630    &    G5Vv    &    0.66    &    5696    &    6.50    &    9.24    &    a    \\
  HD25680    &    G5V    &    0.62    &    5867    &    6.44    &    9.1    &    l    \\
  HD26756    &    G5V    &    0.706    &    5617    &    6.50    &    9.5    &    j    \\
  HD26913    &    G8V    &    0.66    &    5616    &    6.55    &    7.15    &    a    \\
  HD27406    &    G0V    &    0.572    &    6200    &    6.79    &    5.44    &    g    \\
  HD28034    &    G0    &    0.555    &    6146    &    6.65    &    5.2    &    j    \\
  HD28068    &    G1V    &    0.62    &    5758    &    6.64    &    7.73    &    g    \\
  HD28205    &    G0    &    0.545    &    6244    &    6.58    &    6.7    &    j    \\
  HD28344    &    G2V    &    0.619    &    5898    &    6.57    &    7.41    &    g    \\
  HD28805    &    G5    &    0.755    &    5480    &    6.43    &    9.04    &    g    \\
  HD30495    &    G1.5VCH-0.5    &    0.64    &    5836    &    6.51    &    11    &    b    \\
  HD72905    &    G1.5Vb    &    0.58    &    5863    &    6.72    &    4.69    &    a    \\
  HD73350    &    G5V    &    0.669    &    5779    &    6.44    &    6.14    &    z    \\
  HD76151    &    G3V    &    0.67    &    5762    &    6.16    &    15    &    b    \\
  HD95128    &    G1V    &    0.62    &    5869    &    5.74    &    22.7    &    k    \\
  HD97334    &    G0V    &    0.61    &    5850    &    6.64    &    7.6    &    p    \\
  HD101501    &    G8V    &    0.73    &    5450    &    6.29    &    16.68    &    a    \\
  HD103095    &    G8Vp    &    0.75    &    5055    &    5.61    &    31    &    b    \\
  HD114710    &    G0V    &    0.58    &    5963    &    6.34    &    12.35    &    a    \\
  HD115383    &    G0V    &    0.59    &    6009    &    6.66    &    3.33    &    a    \\
  HD115617    &    G7V    &    0.70    &    5558    &    5.81    &    29    &    b    \\
  HD116956    &    G9V    &    0.81    &    5355    &    6.48    &    7.8    &    z    \\
  HD117176    &    G5V    &    0.71    &    5530    &    5.55    &    31    &    s    \\
  HD128987    &    G8Vk:    &    0.737    &    5557    &    6.49    &    9.35    &    z    \\
  HD129333    &    G1.5V    &    0.59    &    5845    &    6.89    &    2.8    &    a    \\
  HD142361    &    G3V    &    0.64    &    5175    &    6.90    &    1.06    &    h    \\
  HD143761    &    G0V    &    0.60    &    5806    &    5.84    &    19    &    s    \\
  HD152391    &    G8.5Vk:    &    0.76    &    5467    &    6.54    &    11.43    &    a    \\
  HD182572    &    G8IV...    &    0.77    &    5595    &    5.85    &    41    &    b    \\
  HD217014    &    G2.5IVa    &    0.70    &    5760    &    5.56    &    37    &    b    \\
  HD283572    &    G5IV    &    0.81    &    4949    &    6.65    &    1.55    &    n    \\
 HD131156A    &    G8V    &    0.777    &    5465    &    6.45    &    6.16    &    l    \\
 \hline
 \end{tabular}
\end{center}
\footnotesize{ \textbf{References:}
(a) \cite{1996ApJ...466..384D}; 
(b) \cite{1996ApJ...457L..99B}; 
% (f) \cite{1985ApJ...294..310B};  % not used
(g) \cite{2003A&A...397..147P}; 
(h) \cite{2006PASP..118.1690M}; 
% (i) \cite{1985AJ.....90.2103S};  % not used
(j) \cite{1984PASP...96..707D};
 (k) \cite{2010MNRAS.408.1666S};
(l) \cite{1987ApJ...316..434S};
%(m) \cite{2010MNRAS.408..475H}; % not used
 (n) \cite{2009A&ARv..17..251S};
% (o) \cite{1995A&A...294..515H}; % not used
 (p) \cite{1984ApJ...279..763N};  
 (r) \cite{2001A&A...371.1024M};
 (s) \cite{2000ApJ...531..415H};
% (t) \cite{1989MmSAI..60..111C};   % not used
 %(u) \cite{1993PASP..105.1407P};  % not used
%(v) \cite{2001ApJ...561.1095B};     % not used 
%(x) \cite{1987ApJ...321..459R};      % not used 
(z) \cite{2000AJ....120.1006G}.\\ 
\textbf{Notes:} Flux is in units of [${\rm{erg}} \cdot {\rm{s}}^{ - 1} {\rm{cm}}^{ - 2} $].}
%\label{default}
\end{table}%

\section{Measuring the $F_{\rm{h+k}}$}   %============================================================

%Method and errors\\
%The Sun\\
%Method justification\\ation based on a local continuum defines by the

For the full stellar sample, we have adopted the process described in  \cite{1984ApJ...279..778H} to measure the observed \ion{Mg}{2} h and k fluxes $f_{\rm{h+k}}$ as the flux integral between the  two minima in the h ($h1v, h1r$) and k ($k1v, k1r$) lines  from the zero flux level  (see Fig. 1). This process was conducted automatically, but verified (and corrected if needed) visually, in particular for stars with lower signal-to-noise in these minima. This method differs from those implemented by \cite{1974A&A....33..257B}, \cite{1978ApJ...220..619L}, and CC07 in that these works either include a correction for the photospheric contribution or perform the integration by considering a local continuum defined by the $k$1 and $h$1 minima. An alternative way to carry out the integration was provided by \cite{2008A&A...483..903B} who considered a fixed 1.7\phantom{a}\AA-width windows to calculate the flux in the k and h lines. In an ideal scenario one should eliminate the photospheric contribution by properly modeling the atmospheric absorption. Nevertheless, the space ultraviolet and in particular the \ion{Mg}{2} features in the mid-UV still represent a  challenge when modeling stellar atmospheres \citep[see, e.g.,][]{2007ApJ...657.1046C}. Additionally, it was argued by  \cite{1984ApJ...279..778H} that the photospheric contribution in the \ion{Mg}{2} lines is much less than for \ion{Ca}{2}. 
Many objects have numerous spectra available, as many as 120 for the case of the star HD2151. For these objects we obtained the mean spectrum by weighting the available spectra by their quoted flux errors.

% HERE was figure 1

\begin{figure}[h!]
\epsscale{1.08}
\plotone{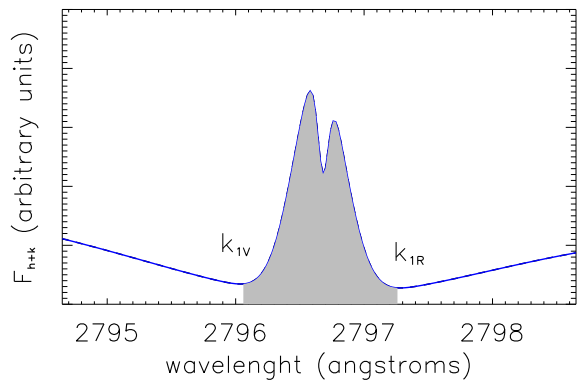}
\caption{Synthetic profile of \ion{Mg}{2} k line. The flux integration is made through the $k1v$ and $k1r$ minima limits.
 \label{fig:MgIIdefinitions}}
\end{figure}

Observed integrated fluxes were then converted to absolute fluxes, $F_{h+k}$, following the 
relation of \cite{1982A&A...110...30O}:
\begin{equation}
\log (F_{\rm{h+k}} /f_{h + k} ) = 0.328 + 4\log T_{{\rm{eff}}}  + 0.4(V + BC).
\label{eq:orange}
\end{equation}
For each object we collected the $V$ magnitudes from SIMBAD database, while the bolometric corrections were taken from \cite{1996ApJ...469..355F}. Effective temperatures are mainly from \citet{2010A&A...515A.111S} and, for a few objects, from \cite{2006ApJ...638.1004A}. In the case of Sun, we have used the IUE spectrum of the Moon LWR09968HS, which was selected among the 21 available spectra, because of its high quality. For the purposes of treating this spectrum in a similar way as the rest of the stars, we have calculated $F_{\rm{h+k}}$ by adopting the V magnitude of the solar analog HD102365, a G2V star with stellar parameters very close to solar: 
($T_{\rm eff}$/log $g$/[Fe/H])=(5637/4.45/-0.08) \citep{1996A&A...314..191G}.
To obtain the error in the surface flux, we applied a Gaussian error propagation through eq.~\ref{eq:orange}. For the effective temperature, we have used the errors provided in \citet{2010A&A...515A.111S} and \cite{2006ApJ...638.1004A} catalogs; for stars with no quoted error in the catalogs, we adopted the mean uncertainty (95 K) of stars with temperature errors available. We assumed for $V$ and $BC$ the conservative values of 1\% and 10\% relative errors, respectively. The uncertainty on the observed flux $f_{\rm{h+k}}$ is obtained from the IUE and HST error vectors as a function of the wavelength; we added in quadrature the mean error in the intervals used to integrate each pair of \ion{Mg}{2} lines. The latter is the main source of error on $F_{\rm{h+k}}$.

\begin{figure}[h!]
\epsscale{1}
\plotone{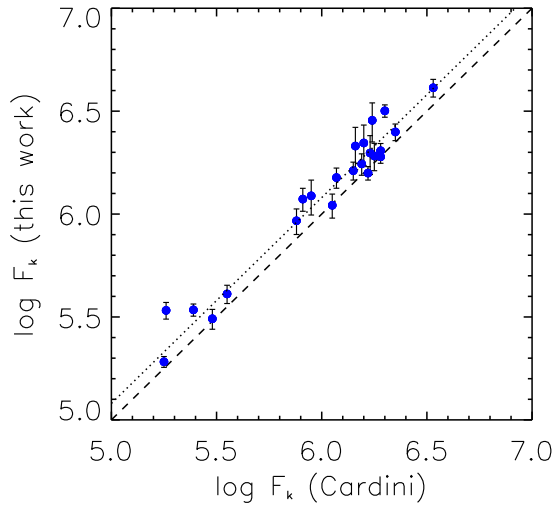}
\caption{The absolute flux for the \ion{Mg}{2} k line ($F_{\rm{k}}$) measured in this work  vs.~those of \cite{2005A&A...430..303C}. The dashed line is the one-to-one relation and the dotted line is a linear fit to the data. A similar procedure is used for the \ion{Mg}{2} h line.
\label{fig:fhkfhk}}
\end{figure}

In order to test the consistency of our derived fluxes we compare in Fig. \ref{fig:fhkfhk} our measurements of the \ion{Mg}{2} k line flux with those reported by \cite{2005A&A...430..303C} (for the 23 objects in common). Whilst the correlation is clear, our fluxes are higher by approximately 25\% (about 0.1 dex) on average. This excess can be plausibly explained by the different integration processes. Since \cite{2005A&A...430..303C} did not report the IUE images that were used, it is possible that they not coincide with ours, as can be the case of HD2151 for which our flux is 80\% larger.

% HERE was figure 2

\section{The $F_{\rm{h+k}}-P_{\rm{rot}}$ Correlation} %===================================================

\begin{figure*}[ht!]
\epsscale{1.5}
\plotone{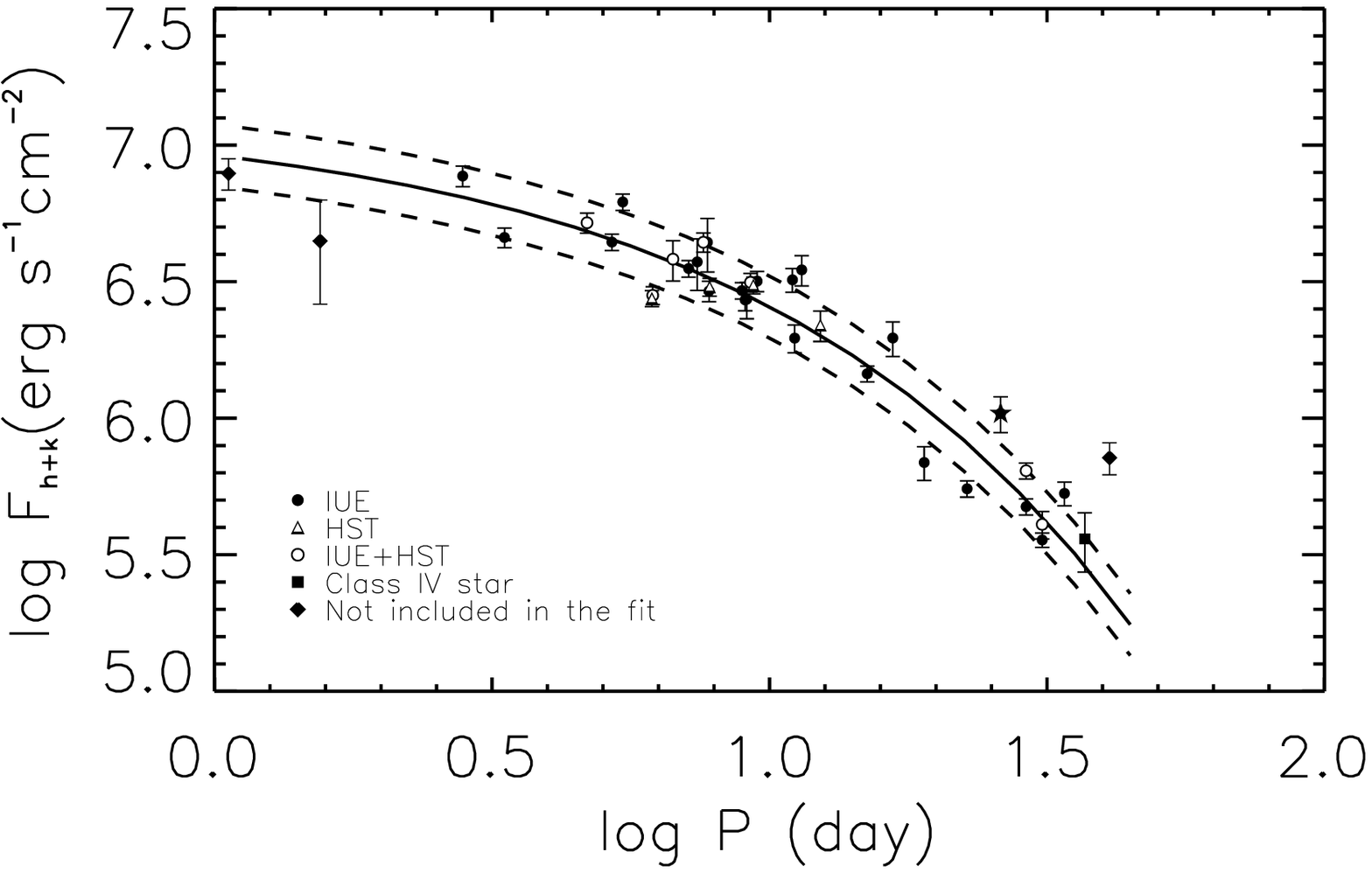} \vspace{-28pt}
\plotone{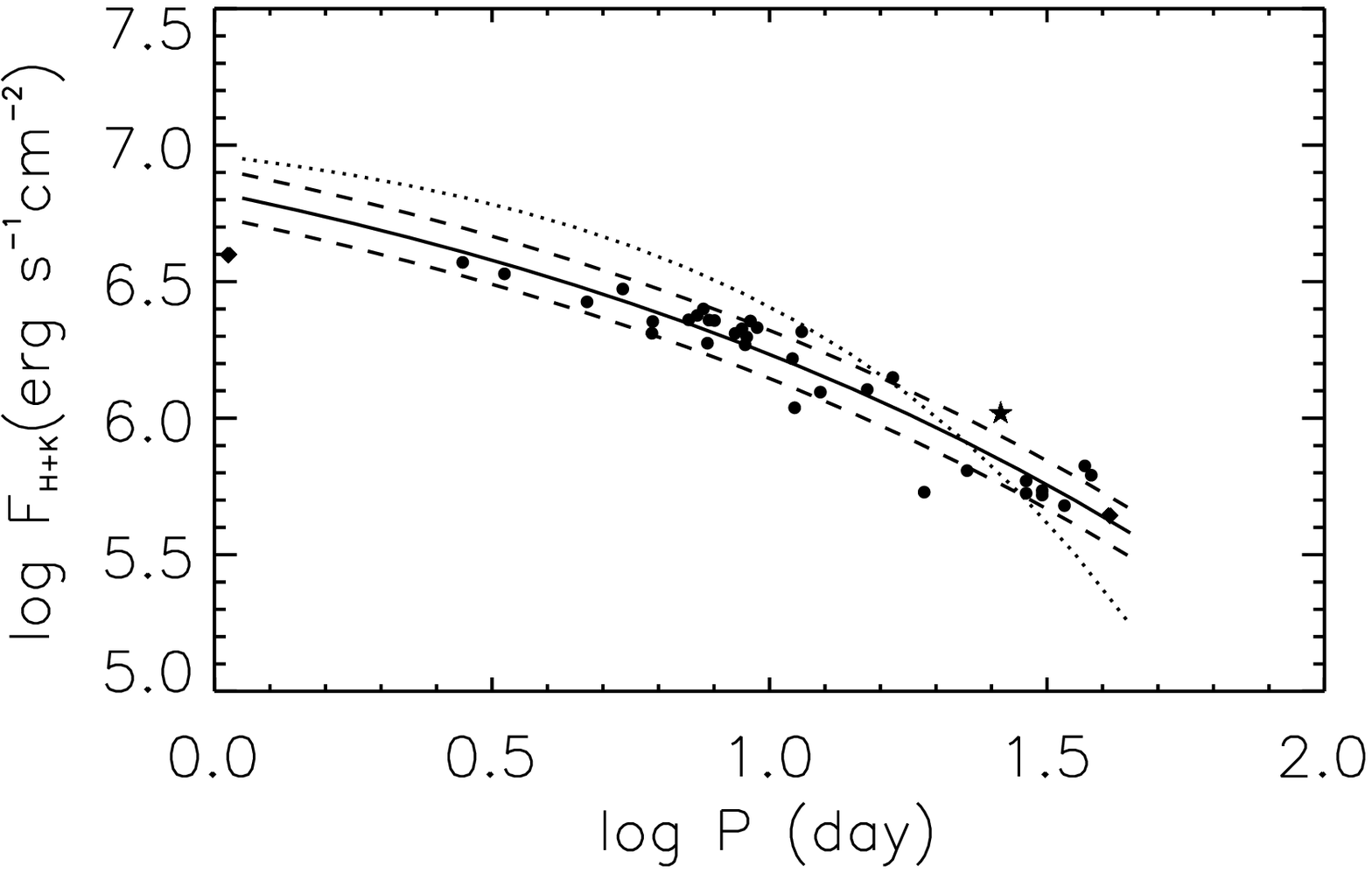}
\caption{Top panel: Mean \ion{Mg}{2} h+k flux ($F_{h+k}$) as function of rotational period. Solid circles, open circles and open triangles are mean $F_{\rm{h+k}}$ from IUE, HST, and  HTS+IUE spectra, respectively. The sun is indicated with a star. The solid line is the best fit to the data, while the dashed lines indicate the $\pm$1 sigma error. Diamonds are stars not used in the fit. Bottom Panel: \ion{Ca}{2} H+K flux vs.~rotational period, $F_{\rm{H+K}}$ derived from the $R_{\rm{H+K}}$ values given by \cite{1996AJ....111..439H} and \cite{2004ApJS..152..261W}. The solid line is the best fit using a similar exponential function as for UV data, shown as a dotted line.
 \label{fig:fhkprot}}
\end{figure*}

In the upper panel of Fig. \ref{fig:fhkprot} we plot $F_{\rm{h+k}}$ vs. the rotational periods 
for the 38 objects (including the Sun)  which have primary values for their rotational periods (Table 1). 
In this plot the different symbols stand for the origin of the data as indicated in the panel. All stars correspond to luminosity classes objects V except for the squared symbol that shows the location of the only luminosity class IV star. The starred symbol indicates the position of the Sun. We have attempted several functional 
forms to calculate the best fit and found that the function:
\begin{equation}
 \log F_{{\rm{h + k}}}  = A + Be^{\log P_{{\rm{rot}}} /C}.
 \label{eq:fit}
\end{equation}     % 7.12530    -0.162499     0.673767
where the coefficient $A=7.125, B=-0.162 , C=0.674$ provided the lowest $\chi^{2}$. The function is 
compatible with the exponential used in \cite{1984ApJ...279..763N}. However, as in \cite{1984ApJ...279..763N}, the adopted functional form does not have  physical implications. 

There is a relatively tight correlation for our sample of G-type stars, with much less dispersion than that observed in other works \citep[e.g.][]{1984ApJ...279..778H, 2007ApJ...666..393C}, that can be explained by the exclusion of the more active cool K and M-type stars. The lines in the plot correspond to the best fitted exponential function (solid line) and the $\pm1\sigma$ levels (dashed lines). 

 % HERE was figure 3ab
 %\begin{multicols}{2}
 
Whilst at low rotational periods ($P_{\rm{rot}} < 4$d ) the diagram is less populated, the correlation indicates that for $P_{\rm{rot}}=0$ the flux appears to reach a maximum of log $F_{\rm{h+k}}\sim6.8$. For periods larger than 4 days there is a decline of more than one decade. The three stars indicated with diamonds were not taken into account in the fitting calculation. The two with the lowest rotational periods correspond to pre-main sequence objects, whose inclusion would produce, in any case, negligible changes in the fitted function. The other object, which shows the largest deviation from the correlation depicted in Fig. \ref{fig:fhkprot}, is the metal-rich star HD182572. 
 This a G8IV variable star whose rotational period of 41 days \citep[$ \log P_{\rm{rot}} = 1.61,$][]{1996ApJ...457L..99B} appears to be twice as large as predicted by its \ion{Mg}{2} flux. Its measured rotational period is compatible with the period of $\sim$37  days estimated from the average rotational velocity ($v \sin{i}$) of 1.9~km\,s$^{-1}$ in \cite{2005yCat.3244....0G} and the stellar radius of $1.38 R_{\sun}$
\citep{2013ApJ...771...40B}. Variability can be a possible explanation for the discrepancy, however, \cite{1997ApJ...485..789L} found the root mean square of the brightness variation in Str\"omgren $b$ and $y$ bands to be just 0.0016~mag during a 12-year long interval (1984-1995). It is possible that an enhanced activity epoch around the IUE observation date (December 1979) produced such a strong UV flux, although this scenario might not be supported by the \ion{Ca}{2} S-index monitoring by \cite{1991ApJS...76..383D}, unless the IUE observation actually coincided with a transient event.

An important aspect of the observed correlation of \ion{Mg}{2} vs. $P_{\rm{rot}}$ is the range of variation of  log$F_{\rm{h+k}}$. The difference between the highest and lowest values is 1.60 dex. In order to compare this difference with that to be expected from \ion{Ca}{2} measurements, we have used the available $R_{\rm{H+K}}$ fluxes for the calcium line from \cite{1996AJ....111..439H} and \cite{2004ApJS..152..261W} for the stars of our sample and transformed them into $F_{\rm{H+K}}$ by multiplying $R_{\rm{H+K}}$ by the bolometric luminosity $\sigma T^{4}$, according to the definition of \cite{1984ApJ...279..763N}. We fit a similar function to eq. \ref{eq:fit} to the  $F_{\rm{H+K}}-P_{\rm{rot}}$ values; this is shown in the lower panel of Fig. \ref{fig:fhkprot}, where the dotted line is the fit for \ion{Mg}{2} case. Note that the variation of $F_{\rm{H+K}}$ is 1.20 dex, i.e. about 0.4 dex smaller than that for \ion{Mg}{2}. We would like to remark that, should we have included the star with the lowest rotational period (HD142361) the \ion{Ca}{2} flux variation would have decreased to 1.09 dex due to the steeper correlation in the low period edge.

Another interesting feature of the panels depicted in fig. \ref{fig:fhkprot} is that the average standard deviation for both correlations are quite similar, with that of \ion{Mg}{2} ($\sigma_{\rm h+k}\sim0.11$ dex) slightly larger than for \ion{Ca}{2} ($\sim0.09$ dex). In this regard, we have conducted a simple test to understand the extent to which the standard deviation can be ascribed to the $F_{\rm{h+k}}$ intrinsic variability of each star. For this purpose, we have compared the amplitude of the variation of the flux in the \ion{Mg}{2} lines, $\Delta \log F_{{\rm{h + k}}}$, with $\sigma_{\rm h+k}$, for the 17 stars with multiple observations. The results of such a comparison indicate that $\Delta \log F_{{\rm{h + k}}}<\sigma_{\rm h+k}$ in 15 out of 17 cases.  The main results of the analysis presented above demonstrates that the magnesium index is about 2.5 times more sensitive to rotational period than that of calcium. \\

\section{New Rotational Periods} %===================================================

\begin{figure}[h!]
\epsscale{1}
\plotone{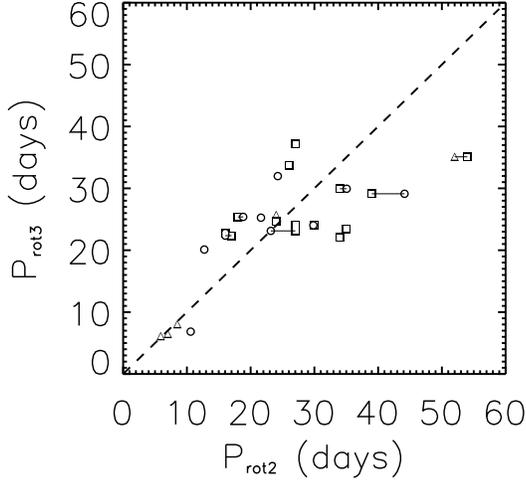}
\caption{Comparison of calculated rotational periods ($P_{\rm{rot3}}$)  and secondary rotational periods from literature ($P_{\rm{rot2}}$). 
Squares and circles have rotational periods from \cite{2004ApJS..152..261W} and \cite{1985AJ.....90.2103S}, respectively, triangles are from \cite{1996ApJ...457L..99B} or  \cite{1997MNRAS.284..803S}. Stars with periods from both groups are connected.
 \label{fig:protlit}}
\end{figure}

A direct application of the exponential fit obtained in the previous section is the determination of new rotational periods ($P_{\rm{rot3}}$) for the objects in Table 3. It is, however, important to compare UV derived periods with those determined from secondary methods, mainly through the calibration of the \ion{Ca}{2} H and K lines. For this purpose we use the second stellar set in Table 2, for which we have collected calculated rotational periods from the literature: these secondary rotational periods ($P_{\rm{rot2}}$)  are mainly from  \cite{1985AJ.....90.2103S},  \cite{2004ApJS..152..261W}, and a few from \cite{1996ApJ...457L..99B} and \cite{1997MNRAS.284..803S}. In Fig. \ref{fig:protlit} we illustrate the comparison of the \ion{Mg}{2} vs \ion{Ca}{2} rotational periods. The dashed line correspond to slope unity. 

In general the points cluster around the one-to-one correlation with less agreement at large rotations periods ($P_{\rm{rot}}>20$d), hence for more mature stars with less intense chromospheric emission. The point that most deviate from the correlation is that for star HD188512 ($\beta Aql$, for which derive $P_{\rm{rot}}=35.1$d significantly smaller than the \ion{Ca}{2} derived periods in excess of 50 days. This star is a subgiant star which has 15 determinations of atmospheric parameters in the compilation of \cite{2010A&A...515A.111S}. The effective temperatures for this object range from $4373 \phantom{a}\rm{K}$ to $5478\phantom{a}\rm{K}$, surface gravities from 1.3 dex to 3.79 dex and a metallicity whose determinations vary nearly 0.6 dex. Even if we exclude the lowest gravity and highest temperature values \citep{2006AJ....131.3069L,1981ApJ...245.1018L}\footnote{We would like to point out that there is a probable identification mismatch. According to \cite{2010A&A...515A.111S}, the lowest value of $\log g$ for HD188512 has been compiled from \cite{1981ApJ...245.1018L}, however, this latter study is for supergiant stars, and the one with the reported parameters actually is $\beta$ Aqr, a GOIb star.}, this star has averaged parameters that are discrepant with the rest of the luminosity class IV stars in our sample, in particular for gravity. Being a more evolved star, the correlation given by eq. \ref{eq:fit} is most probably not applicable to this object.

For the stars in Table 3 we provide for the first time an estimation of their rotational period (column 6) and their uncertainties (column 7). Errors have been obtained through Gaussian propagation of errors of the fitting parameters of eq.~\ref{eq:fit}. Rotational periods for this sample span a wide range of values, from $P_{\rm{rot}}=2.7$d for HD212697 to 35.8d for HD212330. There are several objects with $P_{\rm{rot}}$ very similar to that of the Sun. 

\begin{table}[ht!]
\caption{Basic data for stars with secondary rotational periods}
\vspace{-12pt}
\begin{center}
\footnotesize
\tabcolsep=0.06 cm
\begin{tabular}{llllcllll}
\hline \hline \rule{0pt}{2.5ex} 
Star &Spectral & $B-V$ & $T_{\rm{eff}}$ & $ \log F_{\rm{h+k}}$ & $P_{\rm{rot2}}$ & Ref. & $P_{\rm{rot3}}$ &$ \sigma_{P_{\rm{rot3}}}$   \\ 
&Type&&\scriptsize{[K]}&&\scriptsize{[days]}&&\scriptsize{[days]}&\scriptsize{[days]} \\
[0.5ex]
 \hline \rule{0pt}{2.5ex}
HD26923	&	G0IV	&	0.56	&	5986	&	6.58	&	7	&	c	&	6.5	&	1.7	\\
HD30649	&	G1V-VI	&	0.58	&	5791	&	5.92	&	17	&	w	&	22.4	&	1.7	\\
HD34411	&	G1.5IV-V	&	0.64	&	5821	&	5.61	&	24	&	i	&	32.0	&	1.8	\\
HD43587	&	G0V	&	0.59	&	5899	&	5.82	&	22	&	i	&	25.3	&	1.7	\\
HD52711	&	G4V	&	0.60	&	5891	&	5.82	&	18	&	w	&	25.4	&	1.7	\\
HD67228	&	G1IV	&	0.60	&	5738	&	5.45	&	27	&	w	&	37.2	&	1.8	\\
HD84737	&	G0.5Va	&	0.61	&	5894	&	5.89	&	27	&	w	&	23.1	&	1.8	\\
HD101563	&	G0V	&	0.66	&	5902	&	5.86	&	27	&	w	&	24.0	&	1.7	\\
HD102365	&	G2V	&	0.66	&	5575	&	5.81	&	24	&	d	&	25.7	&	1.7	\\
HD104304	&	G8IV	&	0.78	&	5542	&	5.88	&	35	&	w	&	23.5	&	1.7	\\
HD109358	&	G0V	&	0.60	&	5901	&	5.91	&	16	&	w	&	22.8	&	1.8	\\
HD110897	&	G0V	&	0.50	&	5863	&	6.00	&	13	&	i	&	20.1	&	1.7	\\
HD120066	&	G0.5IV-V	&	0.66	&	5923	&	5.56	&	26	&	w	&	33.7	&	1.8	\\
HD144579	&	G8V	&	0.73	&	5301	&	5.67	&	34	&	w	&	29.9	&	1.7	\\
HD146233	&	G2Va	&	0.65	&	5785	&	5.84	&	24	&	w	&	24.7	&	1.7	\\
HD147513	&	G5V	&	0.60	&	5850	&	6.50	&	8.5	&	d	&	8.1	&	1.7	\\
HD150706	&	G3V	&	0.57	&	5918	&	6.56	&	11	&	i	&	6.8	&	1.9	\\
HD157214	&	G0V	&	0.61	&	5675	&	5.86	&	30	&	i	&	24.0	&	1.7	\\
HD157347	&	G5IV	&	0.65	&	5677	&	5.86	&	30	&	w	&	24.1	&	1.7	\\
HD165185	&	G1V	&	0.57	&	5854	&	6.60	&	6	&	d	&	6.1	&	1.7	\\
HD182488	&	G8V	&	0.79	&	5421	&	5.70	&	39	&	w	&	29.1	&	1.8	\\
HD188376	&	G5IV	&	0.75	&	5505	&	5.93	&	34	&	w	&	22.1	&	1.7	\\
HD188512	&	G9.5IV	&	0.85	&	5091	&	5.52	&	54	&	w	&	35.1	&	1.8	\\
 \hline
 \end{tabular}
\end{center}
\footnotesize{\textbf{References:} (c) \cite{1996ApJ...457L..99B};
(d) \cite{1997MNRAS.284..803S}; 
(i) \cite{1985AJ.....90.2103S}; 
(w) \cite{2004ApJS..152..261W}. \\ 
\textbf{Notes:} Flux is in units of [${\rm{erg}} \cdot {\rm{s}}^{ - 1} {\rm{cm}}^{ - 2} $]. }
%\label{default}
\end{table}

\begin{table}[ht!]
\caption{Stars with newly determined rotational periods}
\vspace{-12pt}
\begin{center}
\footnotesize
\tabcolsep=0.06 cm
\begin{tabular}{llllcll}
\hline \hline \rule{0pt}{2.5ex} 
Star & Spectral & $B-V$ & $T_{\rm{eff}}$ & $\log F_{\rm{h+k}}$ & $P_{\rm{rot3}}$ & $ \sigma_{P_{\rm{rot3}}}$ \\
&Type&&\scriptsize{[K]}&\tiny{[${\rm{erg}} \cdot {\rm{s}}^{ - 1} {\rm{cm}}^{ - 2} $]} &\scriptsize{[days]} &\scriptsize{[days]}   \\ [0.5ex]
 \hline \rule{0pt}{2.5ex}
HD2151&	G0V	&	0.62	&	5773	&	5.76	&	27.1	&	1.7	\\
HD16417	&	G1V	&	0.67	&	5797	&	5.82	&	25.2	&	1.7	\\
HD20794	&	G8V	&	0.71	&	5430	&	5.80	&	25.9	&	1.7	\\
HD44594	&	G1.5V	&	0.65	&	5799	&	5.96	&	21.2	&	1.7	\\
HD59967	&	G3V	&	0.64	&	5780	&	6.56	&	7.00	&	1.7	\\
HD76932	&	G2V	&	0.53	&	5849	&	5.94	&	21.9	&	1.7	\\   %G2VFe-1.8CH
HD136352	&	G4V	&	0.63	&	5623	&	5.78	&	26.7	&	1.7	\\
HD181321	&	G2V	&	0.59	&	5816	&	6.63	&	5.7	&	1.7	\\
HD184499	&	G0V	&	0.59	&	5723	&	5.98	&	20.7	&	1.7	\\
HD190248	&	G8IV	&	0.76	&	5544	&	5.61	&	31.9	&	1.7	\\
HD196378	&	G0V	&	0.51	&	6038	&	5.74	&	27.7	&	1.7	\\  %G0VFe-0.8CH
HD203244	&	G5V	&	0.73	&	5525	&	6.50	&	8.1	&	1.7	\\
HD212330	&	G2IV-V	&	0.68	&	5650	&	5.50	&	35.8	&	1.8	\\
HD212697	&	G3V	&	0.71	&	5769	&	6.82	&	2.7	&	1.9	\\
HD225239	&	G2V	&	0.64	&	5593	&	5.90	&	22.9	&	1.7	\\
 \hline
 \end{tabular}
\end{center}
%\label{default}
\end{table}

\clearpage

\section{Summary}

We have measured the flux of the mid-UV emission lines \ion{Mg}{2} h and k for 76 stars G-type stars (including the Sun) in the main sequence or in the subgiant branch. We used the 38 objects with a rotational period obtained from the periodic variability of their light curves to derive an analytical calibration of $F_{\rm{h+k}}$ vs.~$P_{\rm{rot}}$. We found a tight correlation, which benefits from the inclusion of HST observations as they have a significantly better signal-to-noise ratio compared to the IUE spectra. This is the first time that HST UV data are used in the analysis of this relation, which we used to obtain the first estimates of the rotational periods for 15 stars with high quality mid-UV data.

Our results indicate that the \ion{Mg}{2} h and k lines are about 2.5 times more sensitive to the rotational period than the frequently used chromospheric proxy of \ion{Ca}{2}. The comparison of our UV-derived rotational periods with those obtained from similar correlations for \ion{Ca}{2} shows a significant scatter for slow rotator, where perhaps the very faint \ion{Ca}{2} H and K emission might be prone to large uncertainties. 

\vspace{400 pt}

%\appendix{A. Identification of IUE and HST images}   %===================================================

\begin{appendices}

\center{\chapter{APPENDIX A}}

\center{Identification of IUE and HST images}

\begin{table}[ht!]
\caption{Image identification of some of the spectra used in this work. The full list is available upon request from the authors}
\vspace{-12pt}
\begin{center}
\footnotesize
\tabcolsep=0.9 cm
\begin{tabular}{ll}
\hline \hline \rule{0pt}{2.5ex} 
Star & fits name   \\ [0.5ex]
 \hline \rule{0pt}{2.5ex}
SUN 	&	 LWR09968HS\_2790\_2810	\\
HD001835 	&	 LWP21832HL\_2790\_2810	\\
         	&	 LWR11867HL\_2790\_2810	\\
         	&	 LWP21823HL\_2790\_2810	\\
         	&	 LWP14463HL\_2790\_2810	\\
         	&	 LWR11851HL\_2790\_2810	\\
         	&	 LWP24326HL\_2790\_2810	\\
         	&	 LWP22038HL\_2790\_2810	\\
         	&	 LWP22037HL\_2790\_2810	\\
         	&	 LWP22005HL\_2790\_2810	\\
         	&	 LWP22004HL\_2790\_2810	\\
         	&	 LWP23657HL\_2790\_2810	\\
HD009562 	&	 LWP02200HL\_2790\_2810	\\
HD010700 	&	 LWR03702HL\_2790\_2810	\\
         	&	 LWP04912HL\_2790\_2810	\\
         	&	 LWR11587HL\_2790\_2810	\\
         	&	 LWP02224HL\_2790\_2810	\\
         	&	 LWR13817HL\_2790\_2810	\\
         	&	 LWR04958HL\_2790\_2810	\\
HD011131 	&	 LWP03613HL\_2790\_2810	\\
 \hline
 \end{tabular}
\end{center}
%\label{default}
\end{table}%

\end{appendices}

\clearpage

\bibliography{bibliografia} %===================================================

\end{document}